\documentclass[final,5p,times]{elsarticle}
\usepackage{amssymb}
\usepackage{amsmath}
\usepackage{tikz}
\usepackage{graphicx}
\usetikzlibrary{arrows.meta,positioning,calc}

\journal{Nuclear Physics B}

\begin{document}

\begin{frontmatter}



\title{When Does UML Software Modeling Pay Off? \\ A Contingency Model of Industrial Modeling Value}


\author{Kleinner Silva Farias de Olveira} %% Author name
\ead{kleinnerfarias@unisinos.br}
\affiliation{organization={University of Vale do Rio dos Sinos},%Department and Organization
            addressline={Av. Unisinos, 950, Bairro Cristo Rei}, 
            city={São Leopoldo},
          %   postcode={}, 
            state={Rio Grande do Sul},
            country={Brazil}}

%% Abstract
\begin{abstract}
With the emergence of specification-driven development (SDD), software modeling has regained strategic importance as a way to translate human intentions, requirements, and design decisions into precise specifications that can guide both developers and AI-assisted development tools. Despite modeling's renewed relevance in SDD, empirical evidence remains limited on when its benefits outweigh the socio-technical effort required to create, communicate, synchronize, and maintain software models in industrial practice. This study investigates when software modeling generates value in industrial practice by analyzing 405 survey responses collected from software practitioners. The questionnaire combined closed items concerning UML use, benefits, barriers, and improvement mechanisms with open-ended questions about current use, difficulties, and worthwhile contexts. Most respondents associated modeling with integration discussions (90\%) and shared understanding (82\%), but also reported time pressure (85\%) and the effort of creation and maintenance (80\%). Selective modeling, model–code synchronization, and collaborative support emerged as conditions that improve its viability, whereas only 24\% reported daily UML use. Software modeling pays off when its expected contextual value exceeds the socio-technical friction required to create and sustain it. Its value is therefore contingent, not universal.
\end{abstract}


% \begin{highlights}
% \end{highlights}

%% Keywords

\end{frontmatter}



\section{Introduction}
\label{sec-introduction}
%% Labels are used to cross-reference an item using \ref command.

Software design models (e.g., UML class and sequence diagrams) help practitioners abstract complex software systems, communicate design decisions, and coordinate development activities across requirements, architecture, implementation, and testing \citep{OzkayaErata2020,JongelingEtAl2022}. This role has gained renewed importance as interest in specification-driven and AI-agent-oriented development grows. In this context, requirements, use cases, entity models, architectural constraints, and acceptance criteria become persistent contracts that guide code generation, testing, and validation \citep{Martinelli2026}.

Frameworks and tools such as OpenSpec\footnote{OpenSpec: https://openspec.dev/}, GitHub Spec Kit\footnote{GitHub Spec Kit: https://github.com/github/spec-kit}, BMAD Method\footnote{BMAD Method: https://docs.bmad-method.org/}, and the AI Unified Process\footnote{AI Unified Process: https://unifiedprocess.ai/} exemplify this movement toward structured and traceable development workflows \citep{Martinelli2026,OpenSpec2026,GitHubSpecKit2025,BMAD2026}. Recent evidence from open-source projects also indicates that UML remains present in contemporary software repositories, although its forms and purposes have evolved beyond traditional tool-centered modeling \citep{RomeoEtAl2025}. However, the value of modeling depends on more than a notation's expressiveness or a tool's technical capabilities. Models require time to create, specialized knowledge to interpret, and continuous effort to keep them aligned with other development artifacts (e.g., requirements, source code, tests, and parallel architectural views). Industrial evidence shows that manual consistency management, immature modeling practices, tool-integration difficulties, and legacy artifacts can obstruct short development cycles \citep{JongelingEtAl2022}. Moreover, organizational and human factors shape modeler experience, including motivation, collaboration, versioning practices, language complexity, and tool usability \citep{KalantariEtAl2024}. Consequently, the central question is not simply whether software modeling is useful, but under which conditions its expected value justifies the effort required to create and sustain it.

Previous studies have examined modeling challenges, industrial consistency management, modeler experience, and developers' resistance to modeling \citep{OzkayaErata2020,JongelingEtAl2022,KalantariEtAl2024,KalantariLethbridge2023}. Nevertheless, these complementary lines of research do not sufficiently explain how practitioners balance the expected benefits of modeling (e.g., shared understanding, integration support, architectural reasoning, and complexity reduction) against its socio-technical friction (e.g., time pressure, maintenance effort, organizational resistance, and model--code divergence). Understanding this balance is essential to explain why modeling becomes a valuable engineering investment in some contexts but a dispensable or abandoned activity in others. Therefore, we ask: \textit{When does software modeling pay off in industrial practice?} To answer this question, we analyze evidence collected through a repeated cross-sectional survey of software practitioners and develop a contingency model of industrial modeling value. The model theorizes that modeling is more likely to be adopted and sustained when its expected contextual value exceeds the socio-technical friction involved in producing, interpreting, updating, and integrating its artifacts.

\section{Research Method}

To investigate when UML provides enough value to justify its industrial use, we conducted a time-extended cross-sectional survey with software practitioners between June 2013 and July 2023. Surveys are well suited to examining practitioners' perceptions, reported practices, and organizational conditions across heterogeneous settings~\cite{molleri2016survey,molleri2020checklist}. The ten-year window also reduced the risk that the findings would reflect a single project, technology cycle, or temporary organizational practice. It captures recurring population-level patterns among practitioners surveyed at different points~\cite{brady2015repeated,lebo2015effective}.

This study can reveal persistent benefits, barriers, and adoption conditions, but it cannot measure changes in individual opinions or establish temporal causality. The original dataset contained 416 responses~\cite{silva_farias_de_oliveira_2026_23408327}. Four duplicate submissions and seven incomplete responses with valid answers for fewer than 50\% of the 46 core items were removed. The final sample comprised 405 valid responses.  The questionnaire collected participants' education, professional role, software-development experience, and software-modeling experience. It also contained 46 five-point Likert statements covering perceived benefits, limitations, adoption conditions, and organizational context. Participants answered on a scale from ``Strongly agree'' to ``Strongly disagree.'' A binary item recorded whether they used UML in their daily work.

Three open-ended questions asked how organizations used UML, which difficulties affected its use, and when modeling was worth the required effort. These questions allowed participants to explain concrete practices, e.g., using class diagrams to represent domain structures, sequence diagrams to clarify service interactions, or activity diagrams to communicate business processes. They also revealed practical barriers, e.g., outdated diagrams, weak tool support, and the effort required to synchronize models and source code. Combining closed and open questions provided both measurable distributions and contextual explanations~\cite{jick1979mixing}.

Although the questionnaire did not evaluate current specification-driven development tools, it examined the mechanisms these approaches depend on, e.g., expressing design intent before implementation, coordinating work through shared specifications, and maintaining alignment between specifications and code. UML supports these mechanisms through structural and behavioral representations, e.g., class, sequence, and activity diagrams~\cite{omg2017uml}. The study therefore provides evidence about the conditions under which explicit software representations create value, as well as the costs that may limit their use.

The closed-ended responses were summarized through frequencies and percentages while preserving the five original Likert categories. All open responses were analyzed through iterative open coding, triangulating the quantitative distributions with the qualitative categories~\cite{jick1979mixing}. For example, agreement with a statement about communication was interpreted alongside open responses describing the use of UML in design discussions. Likewise, perceptions of maintenance effort were examined with accounts of obsolete models and model--code divergence. This integration produced the constructs and relationships of the proposed contingency model. The model represents an evidence-grounded explanation of how contextual value and socio-technical friction shape the payoff of modeling; it does not establish causal relationships.

\begin{figure*}[!ht]%% placement specifier
\centering%% For centre alignment of image.
\includegraphics[width=\textwidth]{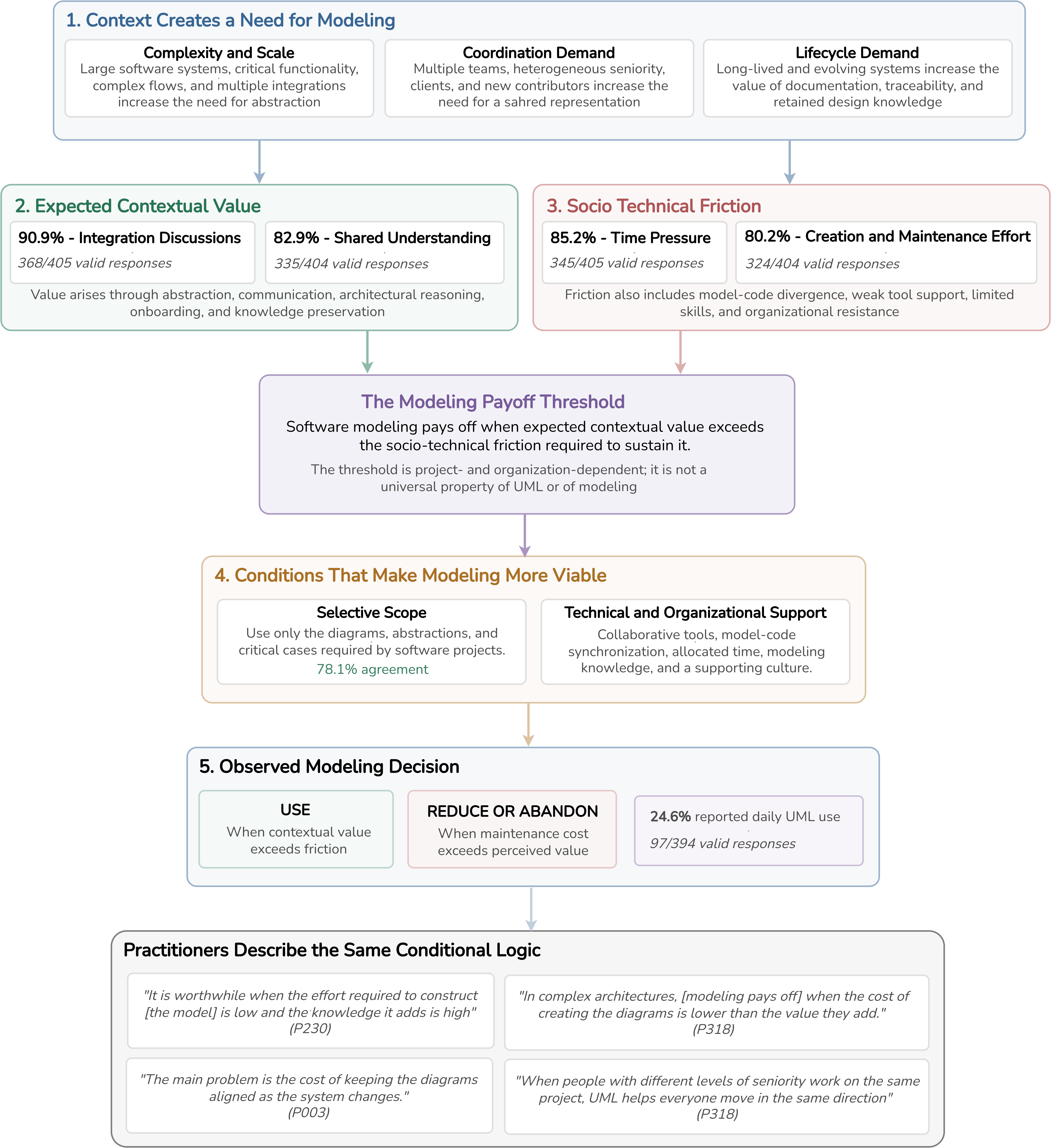}
\caption{Contingency model answering when software modeling pays off in
industrial practice.}
\label{fig:contingency-model-results}
\end{figure*}

%% Use \subsection commands to start a subsection.
\section{Results}
\label{sec-results}

Figure~\ref{fig:contingency-model-results} synthesizes the obtained results as a contingency model of industrial modeling value. The results reveal a marked friction between the perceived value of software modeling and the effort required to maintain it. Regarding expected value, 90.9\% of the respondents agreed that UML models support discussions about application integration (368/405), while 82.9\% associated UML with the development of a shared understanding among developers (335/404). Furthermore, 74.3\% agreed that UML can support communication among professionals with different language abilities (300/404). These results indicate that practitioners value models primarily as instruments for integration, communication, and collective reasoning rather than as ends in themselves. This expected value coexists with socio-technical friction.

Time pressure was recognized as a reason for prioritizing code over models by 85.2\% of the respondents (345/405). Organizational culture affected how UML was used according to 85.6\% (346/404), and 80.2\% reported that creating and maintaining models requires considerable effort (324/404). Difficulties in maintaining consistency across multiple models were also recognized by 65.7\% (264/402). Thus, practitioners may recognize the benefits of modeling while still deciding that its maintenance costs are high in a specific development context.

The responses also suggest two mechanisms to reduce this tension. First, practitioners favor selective modeling: 78.1\% agreed that using only selected diagrams reduces complexity and consistency problems (313/401), 68.6\% reported that models are used mainly for abstractions and critical cases (275/401), and 57.9\% associated modeling value with adapting its use to a specific purpose (232/401). Second, respondents emphasized technical and collaborative enablers. They supported collaborative modeling tools with real-time awareness (84.3\%, 339/402), model--code synchronization (81.9\%, 330/403), feature-oriented diagrams (73.6\%, 296/402), and automatic generation of overview diagrams (64.9\%, 262/404).

The model proposes that modeling pays off when its expected contextual value exceeds the socio-technical friction required to create, interpret, maintain, and integrate its artifacts. Selective modeling reduces this friction, while collaborative and synchronization mechanisms increase the value that teams can extract from models. This interpretation does not imply a causal relationship. It represents an evidence-grounded synthesis of the observed response distributions. The results also reject an indiscriminately negative view of modeling: 60.8\% disagreed that UML fails to meet project needs ($243/400$), 65.9\% disagreed that combining structural and behavioral representations makes systems harder to understand ($265/402$), and 51.5\% disagreed that UML is essentially an academic language ($208/404$). Nevertheless, only 24.6\% of the participants providing an unambiguous answer reported using UML in their daily work ($97/394$), reinforcing the gap between perceived value and routine adoption.

\section{Conclusion}

This study examined when UML provides enough value to justify its use in industrial software development. We analyzed 405 valid survey responses collected from practitioners between June 2013 and July 2023. The questionnaire combined 46 Likert-scale statements with open-ended questions about UML use, difficulties, and conditions of value. We integrated the quantitative distributions with qualitative categories to construct a contingency model. The model relates contextual demands, e.g., architectural complexity, system integration, team coordination, and knowledge preservation, to socio-technical friction, e.g., creation effort, model maintenance, tool limitations, and model--code divergence.

Our findings align with previous studies that highlight the importance of software modeling while rejecting its indiscriminate use. Modeling can support design, documentation, and maintenance comprehension~\cite{torchiano2013relevance}; practitioners also use UML selectively to explore designs, communicate with stakeholders, and describe architectural structures~\cite{junior2022use, petre2013uml}. However, models may become outdated or inconsistent with source code~\cite{junior2022use}, and excessive effort, limited skills, and inadequate tools can hinder adoption~\cite{torchiano2013relevance}. Our study unifies these benefits and barriers in a contingency model: UML pays off when its contextual value exceeds the friction of creating and maintaining it. It contributes a long-term account of industrial perceptions, an evidence-grounded explanation of modeling payoff, and practical guidance for selective modeling. These contributions are especially relevant to specification-driven development, in which explicit representations increasingly guide implementation. Organizations should therefore model with purpose: select the views that support consequential decisions, maintain them while they remain useful, and integrate them with development workflows.

% \begin{figure*}[p]
% \centering
% \resizebox{.97\textwidth}{!}{%
% \input{figure1_layout.tex}%
% }
% \caption{Contingency model about when software modeling pays off in
% industrial practice.}
% \end{figure*}

% %% The Appendices part is started with the command \appendix;
% %% appendix sections are then done as normal sections
% \appendix
% \section{Example Appendix Section}


% %% For citations use: 
% %%       \cite{<label>} ==> [1]

% %%
% Example citation, See \cite{lamport1994document}.

\bibliographystyle{elsarticle-num} 
\bibliography{references}




% %% For numbered reference style

%   \textit{\LaTeX: a document preparation system},
%   Addison Wesley, Massachusetts,

% \end{thebibliography}
\end{document}